\documentclass[english,aps,manuscript,superscriptaddress]{revtex4}
\usepackage{amsfonts}
\usepackage{amsmath}
\usepackage{amssymb}
\usepackage{graphicx}
\usepackage{color}

\providecommand{\U}[1]{\protect\rule{.1in}{.1in}}

\begin{document}

\preprint{}
\title{Discrete Scale Invariance in Topological Semimetals}
\author{Haiwen Liu}
\affiliation{Center for Advanced Quantum Studies, Department of Physics, Beijing Normal
University, Beijing 100875}
\author{Hua Jiang}
\affiliation{College of Physics, Optoelectronics and Energy, Soochow University, Suzhou 215006}
\author{Ziqiang Wang}
\affiliation{Department of Physics, Boston College, Chestnut Hill, Massachusetts 02167 USA}
\author{Robert Joynt}
\affiliation{Kavli Institute of Theoretical Sciences, Chinese Academy of Sciences,
Beijing 100049}
\affiliation{Department of Physics, University of Wisconsin-Madison,
1150 Univ. Ave., Madison WI 53706 USA}
\author{X. C. Xie}
\email{xcxie@pku.edu.cn}
\affiliation{International Center for Quantum Materials, School of Physics, Peking
University, Beijing 100871}
\affiliation{Collaborative Innovation Center of Quantum Matter, Beijing 100871}

\pacs{PACS number}

\begin{abstract}
The discovery of Weyl and Dirac semimetals has produced a number of dramatic
physical effects, including the chiral anomaly and topological Fermi arc
surface states \cite{Wan2011,Vafek2014,Lv2015,Xu2015,Huang2015,Zhang2016a}.
We point out that a very different but no less dramatic physical effect is
also to be found in these materials: discrete scale invariance. This
invariance leads to bound state spectra for Coulomb impurities that repeat
when the binding energy is changed by a fixed factor, reminiscent of fractal
behavior. We show that this effect follows from the peculiar dispersion
relation in Weyl and Dirac semimetals. It is observed when such a material
is placed in very strong magnetic field B: there are oscillations in the
magnetoresistivity somewhat similar to Shubnikov-de Haas oscillations but
with a periodicity in $ln$ B rather than 1/B. \ These oscillations should be
present in other thermodynamic and transport properties. \ The oscillations
have now been seen in 3 topological semimetals: ZrTe$_{5}$, TaAs, and Bi
\cite{Wang2017,Zhang2016b,Behnia2007}.
\end{abstract}

\maketitle

Scale invariance is a prominent ingredient in many important physical
theories, perhaps most notably in the area of phase transitions. \ These
theories contain a field $\phi\left( x\right) $ that has the property that $%
\phi\left( \lambda x\right) =\lambda^{\Delta}\phi\left( x\right) $, where $%
\lambda$ is a continuous parameter and $\Delta$ is the scaling dimension. \
More rare is the occurrence of discrete scale invariance (DSI), where a
similar equation holds, but only if $\lambda$ takes on discrete values \cite%
{Braaten2006}. \ DSI is a characteristic of the fractal behavior that is
seen in many macroscopic systems \cite{Mandelbrot}. \ Its applicability to
bound-state problems in quantum mechanics was exploited by Vitaly Efimov, who
showed that it could occur in the 3-body problem in the limit of resonant
scattering and who derived the energy spectrum in this case \cite%
{Efimov1970, Efimov1973}. \ These ideas have found wide application in
few-body nuclear and atomic physics \cite{Braaten2006, Hammer2010}. \
However, DSI in bound-state problems has generally been considered only for
3- and many-body problems and it has never been experimentally observed in
the solid state. \ In this paper we propose that DSI can occur in the 2-body
problem and indeed that it has been observed in recent experiments in
3-dimensional (3D) Weyl and Dirac semimetals.

The physical systems in question are in high magnetic field and have low
concentrations of mobile electrons such that the ultra quantum limit is
reached, defined by the criterion that the Fermi level is in the lowest
Landau level: $\ell_{B}\lesssim n_{e}^{-1/3}$ where the magnetic length $%
\ell_{B}$ $=\sqrt{\hbar c/eB}$\ and $n_{e}$ is the density of charge
carriers. \ The electrons scatter from impurities, and the temperature is
sufficiently low that the electrical resistivity is entirely due to the
impurity contribution. \ These conditions are readily achieved in modern
experiments \cite{Wang2017,Zhang2016b,Behnia2007}. \ We shall see below from
the experimental phenomenology that the impurities are of the usual types:
short-range scatterers and Coulomb scatterers. \ We focus on the latter type
- indeed it is the Coulomb potential that is the physical source of the DSI.
\ This has dramatic consequences for the magnetoconductivity and other
properties.

It is first necessary to investigate the solutions of the Coulomb problem
for massless Dirac fermions. \ Near any Weyl point in the band structure we
can write the electron Hamiltonian as
\begin{equation*}
H\psi\left( \vec{r}\right) =\left[ v~\vec{\sigma}\cdot\left( \vec{p}-\frac{e%
\vec{A}}{c}\right) -\frac{Ze^{2}}{\kappa r}\right] \psi\left( \vec{r}\right)
=E\psi\left( \vec{r}\right) ,
\end{equation*}
where $\psi\left( \vec{r}\right) $ is a 2-component spinor. \ The
conclusions hold also for the (only slightly more complicated) Dirac case of
a 4-component spinor. \ Here $v$ is the velocity, $\vec{A}$ is the vector
potential, $Ze$ is the charge on the impurity and $\kappa$ is the dielectric
constant. \ In the actual system the $1/r$ potential is cut off at the
atomic scale $r_{a}$ and the fact that it does not diverge is important \cite%
{Pomeranchuk1945}. \

When $\vec{A}=0$ the problem is readily separated \cite{Schiff}. \ The
solutions are
\begin{equation*}
\psi_{jm}^{k}\left( \vec{r}\right) =f_{j}^{k}\left( r\right)
~y_{jm}^{k}\left( \theta,\phi\right) +ig_{j}^{k}\left( r\right)
~y_{jm}^{-k}\left( \theta,\phi\right)
\end{equation*}
Here the $y_{jm}^{k}\left( \theta,\phi\right) $ are the two-component
eigenfunctions of the total angular momentum $j$ with $z$-component $m$ and
the index $k=\ell$ if $j=\ell-1/2$ and $k=-\ell-1$ if $j=\ell+1/2$. \ $k$
effectively labels the parity. \ $\ell$ is the orbital angular momentum. \
We now make the substitution $F\left( r\right) =r~f_{j}^{k}\left( r\right) $
and $G\left( r\right) =r~g_{j}^{k}\left( r\right) $ and drop the indices for
brevity. \ Then we find the radial equations%
\begin{align}
\left( E+\frac{\alpha}{r}\right) F\left( r\right) & =-\partial _{r}G+\frac{k%
}{r}G\left( r\right)  \label{eq:radial} \\
\left( E+\frac{\alpha}{r}\right) G\left( r\right) & =\partial_{r}F+\frac{k}{r%
}F\left( r\right) ~.  \notag
\end{align}
$\alpha=Ze^{2}/\kappa\hbar v$ is the effective fine-structure constant. \
Note that we expect $\alpha\sim O(1)$ since $v<<c.$ \ The character of the
solutions to the radial equations (\ref{eq:radial}) changes abruptly at the
point $\alpha=k$ \cite{Popov1972, Greiner}. \ \ For $\alpha<k$ the solutions
are monotonic in $r$ and cannot show quasi-bound state behavior. \ This is
the subcritical case. \ In the supercritical case $\alpha>k,$ the solutions
are oscillatory and quasi-bound state behavior might be expected. \ The
corresponding problem has been solved in massive 3D systems \cite%
{Popov1972,Greiner} and in massless systems {\color{black} without magnetic field \cite{Shytov2007a, Shytov2007b, Neto2007, Nishida2014, Nishida2016, Zhai2017}.} \ {The Klein tunneling allows the electrons to escape from the attractive potential and in fact only quasi-bound,
non-normalizable states occur.} \ The quasi-energies and widths form a
geometric series. \ Evidence of the 2D supercritical quasi-bound states in
graphene has been seen \cite{Crommie2013, Ovdat2017}.\ \

Equations (\ref{eq:radial}) have the property that if $F\left( r\right) ,G\left(
r\right) $ is a solution with energy $E$ then $F\left( \lambda r\right)
,G\left( \lambda r\right) $ is a solution with energy $\lambda^{\Delta}E$
with $\Delta=1$ for all positive values of the continuous variable $\lambda.$
\ The Efimov case has $\Delta=2.$ \ For bound and quasi-bound states,
boundary conditions break the continuous invariance and give rise to DSI. \
In contrast to the Efimov case we deal here with a single-particle problem
and no parameters in the potential need to take on special values. \ We also
find similarities: in the supercritical regime the wavefunctions have the
log-periodic form $\sin\left[ \frac{1}{2}\sqrt{\alpha^{2}-k^{2}}\ln\left(
r\right) +\delta\right] .$ \ The quasi-bound state energies and widths form
a geometric series (as was also found in 2D). \ These are both general
characteristics of DSI. \ The phase shift $\delta$ is set by the
short-distance cutoff at $r_{a},$ but the details of the short-distance
cutoff are otherwise unimportant. \ The wavefunctions are not normalizable
because the solutions fall off only as $r^{-1}$ and the normalization
integral in 3D diverges at the upper limit.

The main theoretical point of this paper is that an external magnetic field $%
B$ can serve as an effective cutoff at the large distance $l_{B}.$ \ Under
certain conditions to be given below this renders the states normalizable
and suppresses the Klein tunneling, which converts the quasi-bound states to
sharp resonances. \ The binding energy of these resonances depends on the
field in such a way that their energy levels cross the Fermi energy at
discrete values of the magnetic field that constitute a geometric series. \
In fact the form of the wavefunction guarantees that bound or quasi-bound
states will have energies that are periodic in $\ln B$. \ As we will show,
this leads to log-periodic behavior in $B$ of the magneto-conductivity and
other transport and thermodynamic quantities as the quasi-bound state energy
levels pass through the Fermi level. \ This field dependence is reminiscent
of the $1/B$ periodicity that characterizes Shubnikov-de Haas (SdH) and all
other sorts of quantum oscillations. {\color{black}The SdH oscillations in 3D Dirac materials has been systematically studied recently \cite{Shen2018}.} \ Indeed, the log-periodic oscillations
are similar in that they are a consequence of a modulation in the density of
states at the Fermi energy. \ However, $1/B$-periodic quantum oscillations
are intrinsic to solid-state systems. \ The mechanism proposed here is
extrinsic and could be eliminated by purging the system of Coulomb
impurities.

We first present the exact solution for the quasi-energy spectrum and
wavefunctions for $B=0$ for the soft boundary condition $V\left( r\leq
r_{a}\right) =V_{0}=Ze^{2}/\kappa r_{a}.$ \ The main results, as will be
seen, do not depend on the precise form of the boundary condition. \ The
energies and widths are the real and imaginary parts of the poles of the
scattering phase shift as a function of the complex variable $E$. \ The
equation is
\begin{equation*}
\frac{g^{\ast}\left( s_{0}\right) }{g^{\ast}\left( -s_{0}\right) }%
=-e^{-2i\chi\left( -E/\hbar v\right) }\xi~\varepsilon^{\pi s_{0}sgn\left(
E\right) }
\end{equation*}
where $g\left( s_{0}\right) =\Gamma\left( 1+2is_{0}\right) /\Gamma\left(
1+2is_{0}+i\alpha\right) ,$ $s_{0}=\sqrt{\alpha^{2}-k^{2}},$ and $\xi=\left(
\alpha-\sqrt{\alpha^{2}-k^{2}}\right) /k.$ \ $\Gamma$ is the usual gamma
function and the definition of the function $\chi$ in terms of
hypergeometric functions is given in the supplementary information. \ Fig.1
displays the spectrum and wavefunctions. \ They show the characteristic
geometric series behavior $E_{n+1}/E_{n}=e^{-\pi s_{0}}.$\ It is instructive
to compare these to the Wentzel-Kramers-Brillouin (WKB) approximation
combined with the scattering method. The details are given in the supplementary
 informaiton where it is seen that the agreement of the quasiclassical
approximation with the exact solution is very good. \ A very important
consequence of DSI is that the range of validity of the quasiclassical
approximation is very broad and we see that it in fact holds for all $%
r>>r_{a}.$ \ There is no intrinsic length scale at large distances to limit
quasiclassicality. \

We now apply a field $B=B\widehat{z}.$ \ {The magnetic field introduces a
new characteristic length $\ell_{B}$.} \ \ Effectively there is a confining
potential with the characteristic length scale $\ell_{B}$ that now breaks
the DSI at large $r\sim\ell_{B}$. The falloff of the wavefunction with $r$
in the $B=0$ case had the form $r^{-1}$ - this function is normalizable in
1D. \ As $B$ increases, we therefore expect the quasi-bound states to evolve
into true bound states. \ When $B$ is finite, then the problem is no longer
exactly solvable. {\ We consider the case that the expectation vales of the
Coulomb attraction is much greater than the Landau level spacing, which
means that the radius }$r_{n}$ {of the $n$-th quasi-bound state satisfies $%
r_{n}\ll \frac{\sqrt{2}}{2}\alpha l_{B}$. \ In this region the system still
satisfies approximate spherical symmetry, and expanding in spinor spherical
harmonic function reduce the problem to the radial equation, which can be
numerically solved by the WKB method.}

\ We show numerical solutions for the energy spectrum as a function of field
on a \textit{log-log }plot in Fig. 2. \ This is of course not a typical way
to plot an energy level diagram. \ It is appropriate here because of the
very unusual spectrum. \ The quasi-energies increase and the resonances
sharpen as the quasi-bound states as the field increases. \ Only the real
part of the energy spectrum is shown. \ The effective confining potential
raises the levels ultimately into the range of higher Landau levels. \ The
key points about the plot are that at a characteristic field for each level
its energy increases very rapidly and that these characteristic fields also
form a geometric series so that they are equally spaced when plotted as $\ln
B^{1/2},$ as here. \ This is easily understood in the quasiclassical
picture. \ The energy change occurs when $\ell _{B}\approx \hbar
vs_{0}/\left\vert E\right\vert $ and since $E$ forms a geometric series, so
do the characteristic values of $B.$ \ This immediately implies that the
density of electronic states at the Fermi energy is periodic in $\ln B.$ \

The main phenomenological point of this paper is that log-periodic
oscillations have already been seen in 3 semi-metallic systems: ZrTe$_{5},$
Bi$,$ and TaAs \cite{Wang2017,Zhang2016b,Behnia2007}$.$ \ \ In Fig. 3a-3c we
number the peaks and valleys and plot $\ln B$ vs this index at field
strengths that exceed the ultra quantum limit value. \ This type of plot is
a classic way of demonstrating $1/B$-periodic oscillations at lower fields.
\ We see a straight line in the 3 cases where oscillations have been
observed. \ This is very strong evidence for DSI in these materials. \ We
note that the slope, which depends only on $\alpha$ and $k$ varies by a
factor of 2 or so from one material to another. \ Since $\alpha$ depends on
the Fermi velocity and the dielectric constant, this is to be expected. \ \
\

DSI has stronger consequences than simply the $\ln B$ periodicity. \ In fact
we expect that the entire curve of a macroscopic property shows DSI, not
just the peak and valley positions. \ To illustrate this, we compute the
magneto-conductances $\sigma _{xx}~$and $\sigma _{zz},$ since in experiments
at high field it is generally the resistivity tensor derivable from them
that are measured; other properties will be discussed at the end. \

The scattering cross-section for Coulomb impurities at a fixed energy shows
modulation as a function of field: as the bound state energies pass through
a fixed energy the cross section will have a maximum. \ This effect is
captured by the T-matrix approximation for the Kubo formula for $\sigma_{xx}$%
. \ We follow the classic treatment of Bastin \textit{et al}. \cite%
{Nozieres1971}; the details are given in the supplementary information. \ The formula
for $\sigma_{xx}$ is
\begin{equation}
\sigma_{xx}=\frac{4e^{2}}{\hbar}\ell_{B}^{2}\left[ n_{s}+n_{C}\frac{t^{2}}{%
8\pi\hbar v\ell_{\ast}^{-1}\Gamma\left( B\right) }\sum_{n}\frac{\Gamma
^{2}\left( B\right) }{\left[ E_{F}-E_{n}\left( B\right) \right]
^{2}+\Gamma^{2}\left( B\right) }\right] ,  \label{eq:1st form}
\end{equation}
where $n_{s}$ is the density of short-range scatterers, $n_{C}$ is the
density of Coulomb scatters,\ $\ell_{\ast}$ the effective length along the
magnetic field (which depends on microscopic details), $t$ is the coupling
strength between the bound states with the continuum of lowest Landau level
states, $E_{F}$ is the Fermi energy, $E_{n}$ are the quasi-bound state state
energies as calculated and shown above, and $\Gamma\left( B\right) $ are the
widths that are mainly determined by the broadening effect of overlap with
the lowest Landau level. \ The calculations are done at zero temperature. \
This formula in practice is extremely well approximated by \ \
\begin{equation}
\sigma_{xx}=\frac{4e^{2}}{\hbar}\ell_{B}^{2}\left[ n_{s}+n_{C}\frac{t^{2}}{%
8\pi\hbar v\ell_{\ast}^{-1}\Gamma\left( B\right) }\sum_{n}\frac{\eta^{2}}{%
\sin^{2}\left( \frac{s_{0}}{2}\ln\frac{B_{n}}{B_{0}}\right) +\eta^{2}}\right]
,  \label{eq: 2nd form}
\end{equation}
with the fitting parameters are $\eta$ denoting the effective broadening
factor and $B_{0}$ denoting a characteristic magnetic field. Equation (\ref{eq:
2nd form}) is an empirical formula that is convenient for fitting\cite%
{Braaten2006}. \ This formula shows explicitly that $\sigma_{xx}$ peaks at
log-periodic intervals. The longitudinal conductivity along the magnetic
field direction can be also obtained:%
\begin{equation*}
{\sigma_{zz}=\frac{e^{2}}{h}\frac{1}{l_{B}^{4}\cdot16\pi^{2}}\left[
n_{S}+n_{C}\frac{t^{2}}{8\pi\hbar v_{F}\cdot l_{\ast}^{-1}\Gamma\left(
B\right) }\sum_{n}\frac{\Gamma^{2}\left( B\right) }{\left(
E_{F}-E_{n}\left( B\right) \right) ^{2}+\Gamma^{2}\left( B\right) }\right]
^{-1}.}
\end{equation*}
Thus, the longitudinal resistivity also satisfies the DSI property. However,
we expect the DSI feature in $\rho_{zz}$ to be harder to observe than $%
\rho_{xx}$, since the resistivity becomes smaller under large magnetic field
due to the noted negative magnetoresistivity effect \cite{Abrikosov2003}.

Equations (\ref{eq:1st form}) and (\ref{eq: 2nd form}) agree very well; the comparison
is given as follows. {Based on the relation $\sigma_{xy}=\frac{4e^{2}}{h}%
l_{B}^{2}n_{e}$ (}$n_{e}$ {denoting the total carrier density), we obtain
the magneto-resistivity $\rho_{xx}\approx\sigma_{xx}\cdot\sigma_{xy}^{-2}$
(here we used the property $\sigma_{xx}\ll\sigma_{xy}$ from the experimental
data \cite{Wang2017}).} \ In equations (\ref{eq:1st form}) and (\ref{eq: 2nd form}),
the short-range impurity scattering term proportional to $n_{s}$ gives the
well-known linear-B magnetoresistivity \cite{Abrikosov2003}. \ The second
term originates from the resonant scattering process with the quasi-bound
states, which has not been considered previously. \ This new process is
of course what gives rise to oscillations periodic in $\ln B$ in the
magnetoconductances. \ This is seen in the theory and the experiments as an
additional contribution to the linear in B background of $\rho_{xx}.$ \ In
Fig. 3d we show the comparison of the microscopic result in equation (\ref{eq:1st
form}) and the empirical result in equation (\ref{eq: 2nd form}). \ The DSI feature
for the resistivity oscillations become distinct in the log-log plot. The
theoretic log-periodic oscillating curves is used to fit the experimental
data after subtracting the non-oscillating background.

In Fig. 4 we compare our theoretical results to observations in ZrTe$_{5}$
\cite{Wang2017}, using the background subtraction method described in the
supplementary information to remove the the non-oscillating background. \ It is
very important to note that not only do the peaks and valleys show the $\ln B
$ periodicity, but\ indeed the whole curve is log-periodic. \ Our
calculations have been done at zero temperature, which means that at finite
temperatures we need to adjust the broadening parameter $\eta $ to fit
experiment. \ We find that $\eta ^{2}$ equals 0.34, 0.40 and 1.78 at
temperatures 4.2K, 35K, and 80K, respectively. This broadening is due to the
change in the Fermi function. \ The fits are slightly better at small
fields, probably because the shift of the Fermi energy is neglected in the
empirical formula. \ At very high fields the fit is not as good; however,
this has most likely to do with the fact that at the upper limit the
measurement of the field strength becomes less accurate. \ Importantly, the
fitting parameters $B_{0}$ = 0.27 $T$ and $s_{0}$ = 5.2 do not change with
temperature, as theory would, predict. \ Further investigation of the data
reveals that $s_{0}$ does not appear to change from sample to sample. \ This
is significant and since $s_{0}$ has only to do with the characteristics of
the Coulomb potential, one might expect it to be rather universal within one
material. The physical origin of $B_{0}$ is more complicated, since it is
cutoff- and width-dependent. \ It does change for different samples.

The theory may be tested in several ways.

A simple test would be to vary the density $n_{C}$ of Coulomb impurities. \
The amplitude of the oscillations in $\rho_{xx}$ is proportional to $n_{C}.$
\ Varying the valence $Z$ of the impurities would also be of great interest.
\ The ratio of successive peaks in $\sigma_{xx}$ is proportional to $%
\exp\left( -\pi s_{0}\right) $ and $s_{0}=\sqrt{\alpha^{2}-k^{2}}$ with $%
\alpha=Ze^{2}/\kappa\hbar v.$ \ Thus the ratio should decrease as $Z$
increases. \ Another possible effect of increasing $Z\ $is that more than
one angular momentum channel becomes important. \ We have implicitly assumed
above that $k=\pm1.$ \ Then the oscillation pattern would become more
complicated. \ The spectrum of hydrogenic impurities in semiconductors has
been tested in detail by optical experiments. \ This should also be possible
here. \ Since the resonances are only sharp when the field is strong,
field-dependent spectra would be needed.

Our proposal is that the impurities play the central role in creating the
DSI, which means that DSI is an extrinsic effect. \ However, there may also
exist the possibility of an intrinsic effect caused by electron-hole
interactions, \ It may be more difficult to get into the supercritical
regime since $Z$ is fixed at $Z=1,$ but the concentrations would be higher
and the collisions are particularly effective at degrading electrical
current. \ ZrTe$_{5}$ in particular has a temperature-dependent Hall effect
that seems to require both electrons and holes. \

In our theory, we have neglected screening and a possible finite mass, both
effects that break DSI by introducing a length scale. \ The simplest picture
of screening is that the Coulomb interaction is cut off at the Thomas-Fermi
length $L_{TF}=\left( 4\pi e^{2}\partial n/\partial\mu\right)
^{-1/2}=\allowbreak0.32\alpha^{-1/2}g^{-1/6}n^{-1/3},$ where $\mu$ is the
chemical potential and $g$ is the degeneracy including spin and number of
nodes in the band structure. \ This means that $\ell_{B}<L_{TF}$ when the
ultra quantum limit is reached and the $1/r$ form for the potential should
be accurate for our purposes. \ We may also note that screening is expected
to weaken as the field increases. \ We have also assumed zero mass for the
electrons. \ If the mass is not zero, then there is another length $%
L_{M}=\hbar v/E_{g},$ where $E_{g}$ is the energy gap. \ When these lengths
are finite, then DSI holds only for states which have energies deep enough: $%
\left\vert E_{n}\right\vert >s_{0}\hbar v/L_{TF}$ and $\left\vert
E_{n}\right\vert >s_{0}\hbar v/L_{M}.$ \ The oscillations will only be
observed when $L_{M}>>\ell_{B}$. \ This condition puts constraints on the
band structure. \

Comparisons to experiments in 3 materials demonstrate that DSI is a common
feature in topological semimetals. \ It is not a large effect - the amplitude
of the oscillations is at most a few per cent of the total $\rho_{xx}.$ \
But it is a new form of quantum oscillation, dramatically different from
SdH. \ As the theory shows and experiment confirms, it is only expected to
occur in the ultra quantum limit of very high fields and only if the
effective fine structure constant is large enough that the problem is in the
supercritical regime. \ The physics of the log-periodic oscillations in $%
\sigma_{xx}$ is very different from the physics of SdH oscillations.\ \ They
do have one important thing in common, which is that both derive from
oscillations in the density of states at the Fermi energy. \ As a result,
nearly all thermodynamic and transport properties in metals show a $1/B$
periodicity. \ The same should be true of DSI but the oscillations will be
periodic in $\ln B$.


\bigskip

\textbf{Acknowledgements.} We thank Jian Wang, Huichao Wang and Hui Zhai for helpful discussions. This work was financially supported by the National Basic Research Program of China (Grants No. 2017YFA0303301, No. 2015CB921102 and No. 2014CB920901), the National Natural Science Foundation of China under Grants No. 11674028, No. 11534001, No. 11374219, No. 11504008, and NSF of Jiangsu Province, China (Grant No. BK20160007). 


\begin{figure}[tbp]
\includegraphics[width=17cm,height=10.5cm]{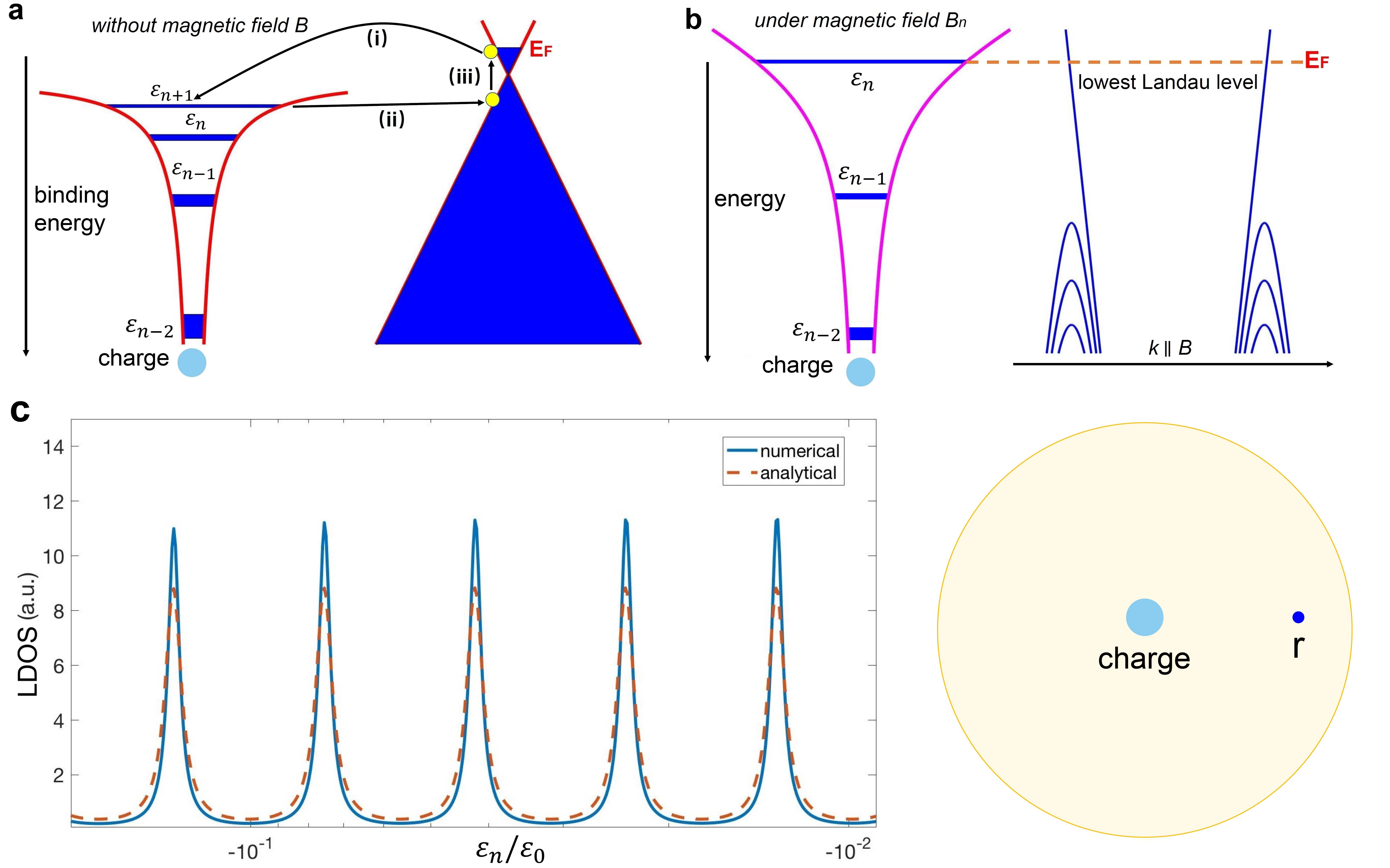} {\color{blue} }
\caption{The DSI properties of quasi-bound states. (a) Without magnetic
field, quasi-bound states form around the charge center [process (i)]. The
Klein tunneling [process (ii)] with electron-hole excitation [process (iii)]
gives rise to finite widths to the quasi-bound state spectra. Due to the
small Fermi surface, screening is relatively weak. \ (b) With a magnetic
field, the potential changes in the large $r$ region, and the quasi-bound
states narrow and move to the Fermi level at certain field values $B_{n}$.
The mobile carriers are in the lowest Landau level when $B$ is in the ultra
quantum limit. (c) The local density of states (LDOS) at an arbitrarily
chosen location with $r>>r_{a}$ as a function of the dimensionless energy
showing the broad quasi-bound states. The exact analytical solutions (solid
line) and the numerical WKB-type solutions (dashed line) are shown. Here $%
\protect\alpha =5.5$ and $k=1$. }
\label{Fig1}
\end{figure}

\begin{figure}[tbp]
\includegraphics[width=16cm, height=8.5cm]{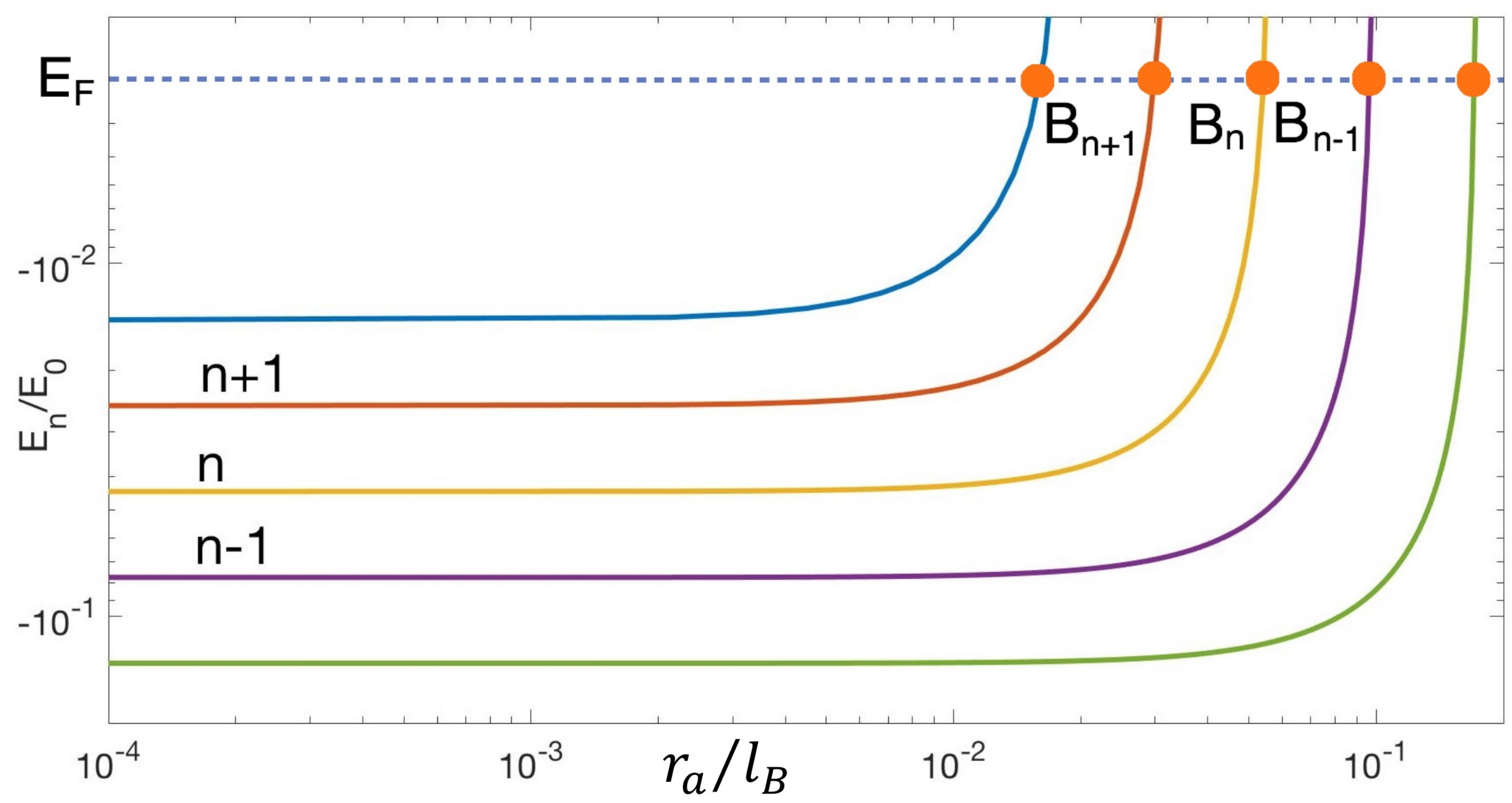}
\caption{The evolution of the dimensionless quasi-energy spectrum versus the
dimensionless inverse magnetic length $\ell _{B}$ for the quasi-bound
states. $r_{a}$ is the short-range cutoff and again $\protect\alpha =5.5$
and $k=1$. Only the real part of five quasi-energies are shown. The broad
quasi-bound states evolve into sharp resonances in the presence of the
magnetic field, and approach the Fermi energy at certain sharp values of the
magnetic field strength $B_{n}$ that form a geometric series.}
\label{Fig2}
\end{figure}

\begin{figure}[tbp]
\includegraphics[width=16cm,  height=16cm]{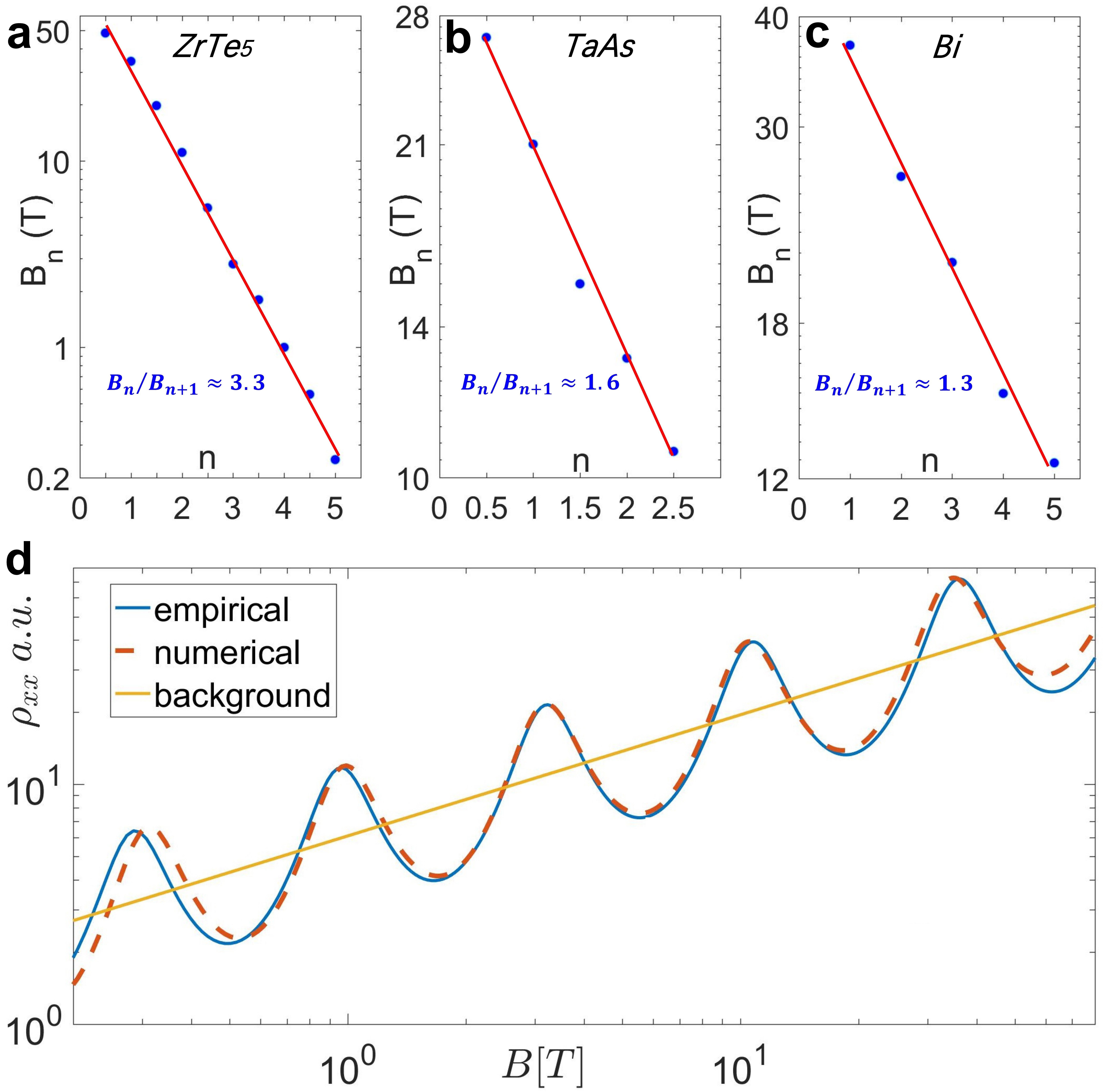} {\color{blue}}
\caption{The DSI features of magneto-oscillations in topological semimetals.
(a)-(c) show that the peaks and dips in magneto-oscillations of ZrTe$_{5}$%
\protect\cite{Wang2017}, TaAs\protect\cite{Zhang2016b} and Bi\protect\cite%
{Behnia2007} form a geometric series. (d) A log-log plot of transverse
resistivity shows oscillations satisfy the DSI property. The empirical
formula (solid line), the microscopic numerical calculation (dashed line),
and the non-oscillating backgrounds are also shown. Here the parameters are $%
\protect\alpha =5.3$ and $k=1$.}
\label{Fig3}
\end{figure}

\begin{figure}[tbp]
\includegraphics[width=16cm, height=20.8cm]{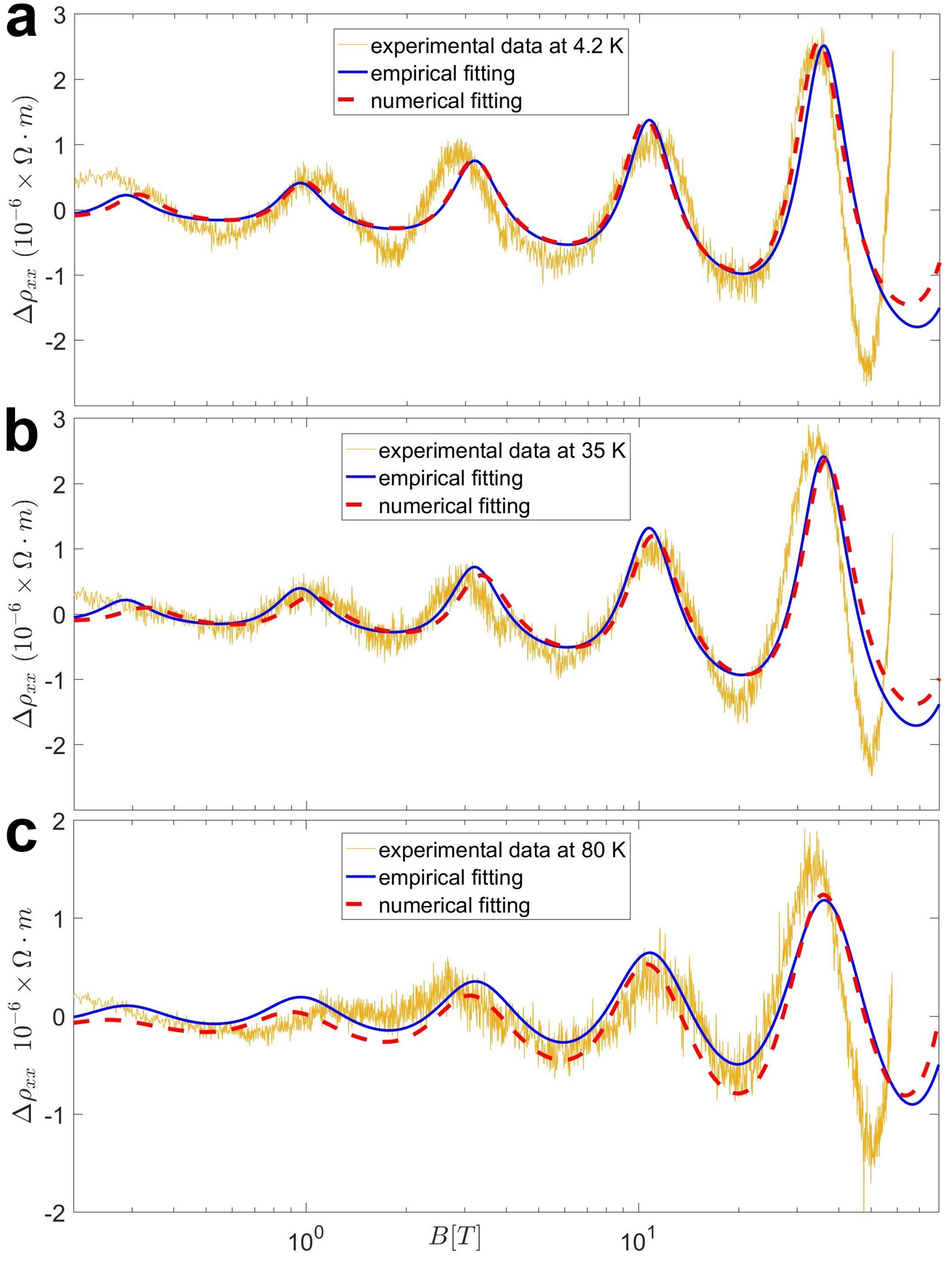}
\caption{The fitting of experimental data for ZrTe$_{5}$ for different temperatures. The resistance data at 4.2K (a), 35K (b) and 80K (c) are shown, with the fitting curves based on the empirical formula (solid lines) and microscopic calculations (dashed lines). The microscopic results are
closer to experimental data at low field, while both empirical and numerical results fit well at high field. The non-oscillating backgrounds are subtracted to fit the experimental data from ref. \protect\cite{Wang2017}.
}
\label{Fig4}
\end{figure}

\end{document}